\documentclass{article}
\usepackage{arxiv}
\usepackage{amsbsy, amstext, amssymb, amsthm, amsmath,bm,bbm,wasysym}
\usepackage{algorithmic,graphicx,booktabs,hyperref,rotating,enumitem,comment}
\usepackage{float}
\usepackage[lined,boxed,linesnumbered]{algorithm2e}
\usepackage{array,stmaryrd}
\usepackage{tabularx}
\usepackage{pdflscape}
\usepackage{adjustbox}
\usepackage{subfig}
\usepackage{longtable}
\usepackage[natbibapa]{apacite}
\usepackage[utf8]{inputenc} % allow utf-8 input
\usepackage[T1]{fontenc}    % use 8-bit T1 fonts
\usepackage{hyperref}       % hyperlinks
\usepackage{url}            % simple URL typesetting
\usepackage{booktabs}       % professional-quality tables
\usepackage{amsfonts}       % blackboard math symbols
\usepackage{nicefrac}       % compact symbols for 1/2, etc.
\usepackage{microtype}      % microtypography
\usepackage{lipsum}
\usepackage{graphicx}

\newtheorem{lemma}{{\bf Lemma}}

\newtheorem{theorem}{{\bf Theorem}}

\newcommand{\Cbf}{{\bm C}}

\newcommand{\greekbold}[1]{\mbox{\boldsymbol{$#1$}}}

\newcommand{\pibf}{\greekbold{\pi}}

\newcommand{\blacktext}[1]{#1}

\title{Probability-scale residuals for event-time data}

\author{
 Eric S. Kawaguchi$^*$ \\
  Department of Population and Public Health Sciences \\
  Keck School of Medicine \\
  University of Southern California\\
  Los Angeles, CA 90032 \\
  \texttt{ekawaguc@usc.edu} \\
   \And
 Bryan E. Shepherd\\
  Department of Biostatistics \\
  Vanderbilt University School of Medicine \\
  Vanderbilt University \\
  Nashville, TN 37235 \\
  \texttt{bryan.shepherd@vanderbilt.edu} \\
  \And
 Chun Li \\
  Department of Population and Public Health Sciences \\
  Keck School of Medicine \\
  University of Southern California\\
  Los Angeles, CA 90032 \\
  \texttt{cli77199@usc.edu} \\
}

\begin{document}

\maketitle

\begin{abstract}
The probability-scale residual (PSR) is defined as $E\{sign(y, Y^*)\}$, where $y$ is the observed outcome and $Y^*$ is a random variable from the fitted distribution. The PSR is particularly useful for ordinal and censored outcomes for which fitted values are not available without additional assumptions. Previous work has defined the PSR for continuous, binary, ordinal, right-censored, and current status outcomes; however, development of the PSR has not yet been considered for data subject to general interval censoring. We develop extensions of the PSR, first to mixed-case interval-censored data, and then to data subject to several types of common censoring schemes. We derive the statistical properties of the PSR and show that our more general PSR encompasses several previously defined PSR for continuous and censored outcomes as special cases. The performance of the residual is illustrated in real data from the Caribbean, Central, and South American Network for HIV Epidemiology. 
\end{abstract}


\section{Introduction}
Residuals are a useful tool for regression analyses.  Besides their traditional roles in evaluation of model fit, residuals can also be used in construction of test statistics \citep{li2010test} and parameter estimators \citep{liu2018covariate}.  The most commonly used residual is the observed-minus-expected residual (OMER), $y-\hat y$, where $y$ is the observed value and $\hat{y}$ is the fitted value.  For ordinal and censored outcomes, fitted values are often not available without making additional assumptions.  Since ordinal regression naturally yields multinomial fitted distributions, \citep{li2010test, li2012new} developed a probability-scale residual (PSR) as a function of the observed value and its corresponding fitted distribution.  
%Specifically, the PSR is $E\{sign(y, Y^*)\}$, where $y$ is the observed outcome and $Y^*$ is a random variable from the fitted distribution, and $sign(a,b)$ is $1$, $0$, $-1$ when $a>b$, $a=b$, $a<b$, respectively.
\cite{shepherd2016probability} further extended the PSR to continuous, binary, and right-censored outcomes, providing a unified framework for residual definition across multiple outcome types.  For right-censored data, the PSR offers a new suite of residuals parallel to martingale, Cox--Snell, and deviance residuals such that each of the three residuals can be expressed as a one-to-one function of the PSR given censoring status \citep{shepherd2016probability}.  The PSR has been a critical component of rank-based measures such as partial Spearman correlation \citep{liu2018covariate, eden2023spearman} and rank intraclass correlation \citep{tu2023rank}.

In addition to right censoring, data collected in research can be subject to other types of censoring.  \blacktext{In early childhood development studies, interest may lie in determining when a child is able to complete a certain task. The age at which a child accomplishes this task is treated as the time-to-event. In some cases, children may already be capable of performing the task when enrolled in the study. These event times are referred to as left censored. While not as common in event-time studies, left censoring is prevalent in environmental studies where lab-recorded measurements are not quantifiable below a certain threshold value (the detection limit). A related censoring mechanism is doubly-censored data where the outcome is subject to both left censoring and right censoring. With event-time data, double censoring occurs when both the time origin and time-to-event are subject to censoring \citep{de1989analysis}. In cancer screening studies, subjects who had a tumor before their first screening are considered left censored, while right censoring occurs when a tumor is not detected by the end of the study.} 

In clinical or epidemiological settings, the event time of interest may or may not be directly observed.  When it is not directly observed, it is only known to have occurred between successive examinations or visits. In such cases, data are said to be interval censored since the time to event is only known to lie within an interval. interval-censored data is often characterized by the number of {\blacktext{examination}} times, denoted as $k$. Case I interval-censored data is often referred to as current status data \citep{huang1996efficient, lin1998additive, martinussen2002efficient}. In this setting, an individual is examined only once ($k = 1$), and the outcome is considered left censored if the event already occurred by the examination time or right censored if the event has not occurred. interval-censored data for $k\ge2$ have been explored by \cite{groeneboom1992information} and \cite{wellner1995interval}, where all individuals are examined exactly $k$ times. Since in many studies it is unlikely that every participant of a study has exactly the same number of visits, a more general and practical setting is 
%In reality, settings with $k \geq 2$ are difficult to find since it is unlikely that every patient under study has exactly the same number of visits. 
%For example, if the event has occurred before the $k$th examination, further examinations become unnecessary.
%A broader framework that encompasses Case $k$ interval censoring is mixed-case interval censoring.
{\blacktext{mixed-case interval censoring, in which $k$ is treated as a random integer rather than a constant }}\citep{schick2000consistency}.  Commonly used regression models for right-censored data have been extended to interval-censored data \citep{finkelstein1986proportional, zeng2006semiparametric, tian2006accelerated}; and several well-known residuals for right-censored data have been extended to interval-censored data \citep{farrington2000residuals}, which have been used to construct goodness-of-fit tests \citep{sakurai2018goodness}.

The various scenarios of event-time data described above are special cases of a general setting in which the event times of some individuals are directly observed and the other individuals are right-, left- or interval-censored \citep{kim2003maximum}.  In this general setting, the outcome is a mixture of censored and uncensored observations.  It is desirable to have a coherent definition of residual for all the scenarios of event-time data.  The unifying nature of PSR for various outcome types --- continuous, ordinal, and censored --- makes it a promising candidate.

In this paper, we first define in Section \ref{sec:method} the PSR for mixed case interval-censored data.  We will describe its properties and connections with other residuals for interval-censored data.  We then develop a ``unified" PSR for the general setting in which the outcome is a mixture of censored (right, left, interval) and uncensored observations.  This covers several commonly-seen censoring schemes in practice. A real data application is presented in Section \ref{sec:rda} using data from the Carribean, Central American, and South America network (CCASAnet) for HIV epidemiology. Lastly, concluding remarks are provided in Section \ref{sec:disc}.

%Interval censoring assumes that no event times are exactly observed. In practice, it is possible that data arise as a mixture of censored (right, left, interval) and uncensored observations. It is of interest, to develop a unifying PSR framework that encompasses these data. Motivated by this need, we first introduce the PSR for mixed case interval-censored data in Section \ref{sec:method}. We describe its statistical properties and explore its relation to the PSR defined in \cite{shepherd2016probability} for continuous, right-censored and current status data.  We then develop a ``unified" PSR for event-time data that captures several commonly-seen censoring schemes in practice and derive its statistical properties. A real data application is presented in Section \ref{sec:rda} using data from the Carribean, Central American, and South America network (CCASAnet) for HIV epidemiology. Lastly, concluding remarks are provided in Section \ref{sec:disc}.

\section{Methods}
\label{sec:method}

\subsection{A brief introduction to probability-scale residuals}

The probability-scale residual (PSR) was first introduced as a single-valued residual for ordinal categorical outcomes \citep{li2012new} and had been generalized to continuous, binary, count, and censored data \citep{shepherd2016probability}.  The PSR is a function of an observed value $y$ and a fitted distribution $F^*$ from a regression model,
\begin{equation}
    \label{eq:psr}
    r(y, F^*) =F^*(y-) + F^*(y) - 1,
\end{equation}
where $F^*(y-) = \lim_{t \uparrow y} F^*(t)$.  The motivation behind the PSR is to contrast the observed value $y$ with a value $Y^*$ randomly drawn from the fitted distribution $F^*$ using $\text{sign}(y, Y^*)$, where $\text{sign}(a, b)$ is $-1$, $0$, and $1$ for $a<b$, $a=b$, and $a > b$, respectively.  The PSR is the expected value of this contrast function with respect to $Y^*\sim F^*$:
\begin{equation*}
    E\{\text{sign}(y, Y^*)\} = \Pr(Y^* < y) - \Pr(Y^* > y) = F^*(y-) + F^*(y) - 1.
\end{equation*}
%Letting $Y$ be an orderable random variable (continuous or discrete) from a distribution $F$, $y$ being an observed value of $Y$ and $F^*$ being an assumed or fitted distribution of $Y$, the PSR is defined as
%\begin{align}
%    \label{eq:psr}
%    r(y, F^*) = E\{sign(y, Y^*)\} = \Pr(Y^* < y) - \Pr(Y^* > y) = F^*(y-) + F^*(y) - 1,
%\end{align}
%where $Y^*$ is a random variable with distribution $F^*$, $F^*(y-) = \lim_{t \uparrow y} F^*(t)$, and $sign(a, b)$ is $-1, 0,$ and $1$ for $a<b, a = b,$ and $a > b$, respectively.
The residual was shown to have important and desirable properties. In particular, the PSR is easy to understand and interpret, is bounded between $-1$ and $1$, does not require a fully specified fitted distribution, and has expectation 0 when $F^*$ is the correct distribution for the observed value $y$. Another advantage of the PSR is its consistent definition across outcome types. This sets it apart from the classical observed-minus-expected residual (OMER), which lacks a clear definition for several outcome types.

%Typically the distribution of $F^*$ will be conditional on covariates $Z$ and corresponding parameters $\thetabf$ denoted as $F_{Z; \thetabf}$. For $i = 1, \ldots, n$ subjects, let $Y_i$ be the outcome of interest and define $F^*_{Z_i; \thetabf}$ as the assumed distribution of $Y_i$ given a set of covariates $Z_i$. Then for given data $(y_i, z_i)$ with estimated parameters $\hat{\thetabf}$, the PSR is given by $\hat{r}_i = r(y_i, F^*_{z_i; \hat{\thetabf}})$. For notational convenience, we will omit covariates.

When $F^*$ is a continuous distribution, the PSR becomes $2F^*(y) - 1$.  In comparison, the classical OMER is $y-E(Y^*)$, where $Y^*\sim F^*$.  In linear regression with a constant residual variance, the fitted distribution may be derived as a normal distribution $N(\hat y, \hat\sigma^2)$, where $\hat y$ is the fitted value, \blacktext{possibly dependent on a set of covariates $Z$}, and $\hat\sigma^2$ is the estimated residual variance.  For this scenario, the OMER is $y-\hat y$ and the PSR is $2\Phi((y-\hat y)/\hat\sigma)-1$, where $\Phi(\cdot)$ is the cumulative distribution function of the standard normal distribution.  \blacktext{It is evident that the PSR is a transformation of the OMER and thus captures the same information as the OMER but delivers it on a different scale.} 

For right-censored data, let $T>0$ be the time to the event of interest, $C\ge0$ be the time to censoring, $Y=\min\{T,C\}$, and $\Delta=I\{T\le C\}$. For each individual, a realization $(y,\delta)$ of the random vector $(Y,\Delta)$ is observed. \blacktext{Here, fitted distributions are fully specified when parametric models are used to fit the data (e.g., parametric accelerated failure time models). When semi-parametric models are employed (e.g., the \cite{cox1972regression} proportional hazards model) the fitted distribution can only be estimated at the observed distinct event times.} When the individual is uncensored (i.e., $\delta=1$), the PSR is defined as in (\ref{eq:psr}). When the individual is right censored (i.e., $\delta=0$), we only know the event time will occur after the censoring time.  In this situation, the PSR is defined as the expectation of $r(T^*,F^*)$ with respect to $T^*\sim F^*$ conditional on $T^*>y$. \cite{shepherd2016probability} showed that the PSR is $F^*(y)$ in this situation, and a more general expression of the PSR that covers both uncensored and censored cases is
$r\{(y,\delta),F^*\} =F^*(y)-\delta(1-F^*(y-))$. \cite{shepherd2016probability} demonstrated that the PSR for right-censored data exhibits analogous properties to their uncensored counterpart.

\subsection{Probability-scale residual for mixed-case interval-censored data}

%Before introducing a general-use PSR for censored data, we first define the PSR for  mixed-case interval-censored data. In the case of
With interval-censored data, we do not directly observe the event time of interest; instead, we only have knowledge that the event time falls within an interval derived from a series of examination (censoring) times. 
%Current status data can be thought of as an extreme case of interval censoring with one examination time.
While there are several different ways to characterize interval-censored data, we adopt the mixed-case interval censoring setup \citep{schick2000consistency} in this paper. 

Let $T>0$ be the time-to-event outcome and $F(t)$ be its cumulative distribution function such that $F(0)=0$ and $F(\infty)=\lim_{t\to\infty}F(t)=1$.  Let $K$ be a random positive integer.  Given $K = k$, we observe a random vector of examination times, 
$\Cbf=(C_1, \ldots, C_k)$, such that $0<C_1<\cdots<C_k<\infty$. The censoring information is determined by $(K,\Cbf)$. For convenience, we let $C_0=0$ and $C_{k+1}=\infty$, and use the notation $(l,u]$ to denote a half-open interval when $0\le l<u<\infty$ and an open interval $(l,\infty)$ when $u$ is infinite. The vector $\Cbf$ divides the space $(0,\infty)$ into $k+1$ intervals. We do not directly observe $T$ and only know that it falls in an interval $T \in (C_j, C_{j+1}]$ for some $j = 0, \ldots, k$. The lower and upper endpoints of this observed interval are $L = C_{j}$ and $U = C_{j+1}$, respectively.
%with $L = \max\{C_j: C_j < T; j = 0, \ldots, k\}$ and $U = \min\{C_j: C_{j} \geq T; j = 0, \ldots, k + 1\}$.
Note that a subject is considered left censored when $L = C_0 = 0$ and right censored when $U = C_{k+1} = \infty$.

If we had observed the event time $T=t$, the PSR would be defined as $r(t,F^*) = F^*(t) + F^*(t-) - 1$, where $F^*$ is the fitted distribution of $T$, \blacktext{which can be estimated parametrically or semi-parametrically \citep[see e.g.,][]{finkelstein1985semiparametric, finkelstein1986proportional, odell1992maximum, rabinowitz1995regression, zhang2005regression, zeng2006semiparametric}}. However, for interval-censored data, we instead only observe an interval $(L,U]=(l,u]$, where $0 \leq l <  u$. Similar to the definition of PSR for right-censored data, we define the PSR for interval-censored data as the expectation of $r(T^*,F^*)$ with respect to $T^*\sim F^*$ conditional on $T^*\in (l, u]$:
\begin{align}
    \label{eq:PSR}
    r\{(l, u], F^*\} & = E\{ r(T^*, F^*) \mid T^* \in (l, u] \} \notag \\ 
    & = F^*(l) + F^*(u) - 1.
\end{align}
The details of the derivation are in the Appendix. %\blacktext{Note that if $T$ is exactly observed, then we can consider the ``interval" $(t-, t]$ in which case the PSR would be simplified to its uncensored form, $r\{(t-, t], F^*\} = F^*(t) + F^*(t-) + 1$. With $t- \to t$, the residual would approach the PSR for continuous outcomes, $2F^*(t) - 1$.}

\subsubsection{Properties of the PSR for interval-censored data}

The PSR defined in (2) for interval-censored data satisfies the following properties:
\begin{enumerate}
    %\item[P1)] For any $a \in (0, \infty)$, $-1 \leq r\{(0, a], F^*\} \leq  r\{(a, \infty), F^*\} \leq 1$.
    \item[P1)] $r\{(0, \infty), F^*\} = 0$.
    \item[P2)] $-1\le r\{(l, u], F^*\} \le1$, for any $0\le l<u$.
    %\item[P2)] For any $0 \leq a < b < \infty$. If $F^*$ is invertible, then $r\{(a, b], F^*\} = -r\{(a^*, b^*], F^*\}$ where $a^* = F^{*-1}[1 - F^*(b)]$ and $b^* = F^{*-1}[1 - F^*(a)]$.
    \item[P3)] Let $0\le l_1 < u_1$ and $0\le l_2 < u_2$ with $l_1 \leq l_2$ and $u_1 \leq u_2$. Then $r\{(l_1, u_1], F^*\} \leq r\{ (l_2, u_2], F^*\}$. 
%For Property 2, $F^*$ can be assumed to be invertibile since it is monotonic. Property 3 allows us to order residuals under certain conditions. As a consequence, one can show that $r\{(0, a], F^*\} \leq  r\{(a, \infty), F^*\}$ for any $a \in (0, \infty)$.
\end{enumerate}
Property P1 shows that the PSR is zero if there is no information about the event time.  Property P2 shows that the PSR is bounded between $-1$ and $1$. Property P3 describes a desirable monotonic ordering of the PSR. When an interval $(l_2,u_2]$ is "on the right" of another interval $(l_1,u_1]$ in the sense that $l_1\le l_2$ and $u_1\le u_2$, the residual of $(l_2,u_2]$ is no lower than that of $(l_1,u_1]$.  A special case is when two intervals share an endpoint: when they have the same upper endpoint $u$ while $l_1<l_2$, then $r\{(l_1, u], F^*\} \leq r\{(l_2, u], F^*\}$; when they have the same lower endpoint $l$
while $u_1<u_2$, then $r\{(l, u_1], F^*\} \leq r\{(l, u_2], F^*\}$.

In addition, when $F^*$ is $F$, the true distribution of $T$, and $T$ is independent of the censoring variables $(K,\Cbf)$, we have
\begin{enumerate}
    \item[P4)] $E[r\{(L, U], F\}] = 0$.
    %\item[P5)] $Var[r\{(l, u], F^*\}] = E_{C, K} F(U)[1 - F(L)^2 - F(L)F(U) - F(U)]$.
    \item[P5)] $Var[r\{(L, U], F\}] = E_{(K, \Cbf)} \left(\sum_{k=1}^K F(C_{k+1})F(C_k)[F(C_{k+1}) - F(C_k)]\right)$.
\end{enumerate}
Their derivation is in the Appendix. %Similar to the variance of the PSR for right-censored outcomes, the 
The variance of the PSR depends on the distributions of both $T$ and $(K,\Cbf)$. We note that the conditional variance of the PSR given $(K,\Cbf)$, $\sum_{k=1}^K F(C_{k+1})F(C_k)[F(C_{k+1}) - F(C_k)]$, has a similar form as the variance of the PSR for ordered categorical outcomes (see property 9 of \cite{li2012new}).  This makes sense because once the values of $K$ and $\Cbf$ are fixed, interval-censored data are effectively ordered categorical data.

\subsubsection{Connection of the PSR with other residuals for interval-censored data}
In the case of right-censored data, various context-specific residuals like the Cox-Snell, martingale, and deviance residuals have gained widespread acceptance as tools for model diagnostics [see e.g. \citep{kay1977proportional, barlow1988residuals, therneau1990martingale}]. The Cox-Snell residuals can be used to assess the overall fit of a model. If the model fits the data well, then the Cox-Snell residuals should follow a unit exponential distribution. This is often assessed by plotting the residuals against the Nelson-Aalen estimator of the residuals. The martingale residuals are a slight modification of the Cox-Snell residuals and have the property of being centered around zero. These residuals are often used to determine the functional form of a covariate. Martingale residuals are highly skewed since they take on values between $(-\infty, 1)$. The deviance residual is a normalized transform of the martingale residual and is ideal for outlier detection. \cite{shepherd2016probability} show that the PSR for right-censored data can be expressed as one-to-one functions of these residuals.

Residuals analogous to the Cox-Snell, martingale, and deviance residuals have been developed for interval-censored data \citep{farrington2000residuals}.  Unlike the PSR derived for right censored outcomes, there is no trivial one-to-one correspondence between the PSR for interval-censored data and the suite of residuals proposed in \citep{farrington2000residuals}. However, there is a connection between the PSR and the Cox-Snell residual for interval-censored data. Due to the nature of the data, the interval censored version of the Cox-Snell residual is not easy to handle since, by definition, they themselves are intervals. To create a single-valued residual as an alternative, the adjusted Cox-Snell residual was suggested by replacing the interval residuals with expected values under the unit exponential distribution. We show, in the Appendix, that by replacing the interval residuals with expected values under the $\mathcal{U}(-1, 1)$ distribution will produce the PSR defined in (\ref{eq:PSR}).

\subsection{A unified PSR for censored data}
\label{sec:upsr}

The mixed-case interval censoring setup described in Section 2.2 assumes that true failure times cannot be exactly known. However, as we will see in Section \ref{sec:rda}, one may have a study where some failure times are exactly known and others are subject to one of several censoring schemes.  %Thus, data for subject $i$ can be represented as $\{\Delta_i, \Delta_i T_i, (1 - \Delta_i)L_i, (1- \Delta_i)U_i\}$, where $\Delta \in \{0, 1\}$ indicates whether $T$ is exactly observed or not and $L_i$ and $U_i$ are defined as in Section \ref{sec:method}. 
It is desirable to have a coherent definition of residual for all the scenarios of event-time data. The unifying nature of PSR for various outcome types makes it a promising candidate. To define the PSR, we now extend the mixed case interval censoring setup to encompass various censoring situations.  Specifically, given $K=k$ and $\Cbf=(C_1,\ldots,C_k)$, suppose there are $k + 1$ mixing probabilities $\pibf =\{\pi_0,\pi_1,\ldots,\pi_k\}$ such that when $T\in(C_j,C_{j+1}]$, with probability $\pi_j$ we observe the exact value of $T$, and with probability $1 - \pi_j$ we observe only the interval $(C_j,C_{j+1}]$. The outcome $y$ is either an observed event time $t$ or an interval $(C_j,C_{j+1}]$ that the unobserved event time falls into.

This general setup covers many censoring schemes commonly seen in practice, as detailed below.
\begin{enumerate}
    \item[S1.] Continuous event-time data with no censoring: $\pi_j\equiv1$ for all $j$;  
    \item[S2.] Right-censored data: $K=1$, %i.e., $C = \{0, C_1, \infty\}$, 
    $\pi_0=1$, and $\pi_1=0$;
    \item[S3.] Left censored data: $K=1$, $\pi_0=0$, and $\pi_1=1$;
    \item[S4.] Doubly censored data: $K=2$, $\pi_0=0$, $\pi_1=1$, and $\pi_2 = 0$;
    \item[S5.] Current status data: $K=1$, $\pi_0=0$, and $\pi_1=0$;
    \item[S6.] interval-censored data: $\pi_j\equiv0$ for all $j$.    
\end{enumerate}

\begin{comment}
To extend the PSR methodology to general time-to-event data, we would need to modify the PSR to allow for situations when $T$ is exactly observed. We do so by decomposing the PSR conditional on $\Delta$:
\begin{align}
    \label{eq:mPSR}
    & r\left(\{\Delta, \Delta T, (1 - \Delta)L, (1- \Delta)U\}, F^* \right) \notag \\
    & = \Delta r(t, F^*) + (1 - \Delta) E\{r(T, F^*)|T \in (l, u]\} \notag \\
    & = \Delta \left\{ F^*(t) + F^*(t-) - 1 \right\} + (1 - \Delta) \left\{ F^*(u) + F^*(l) - 1\right\}.
\end{align}

Thus, the PSR is a mixture consisting of two components, the standard PSR defined by \cite{shepherd2016probability} when $T$ is exactly observed ($\Delta  = 1$) and the PSR when $T$ is only known to be within some interval ($\Delta = 0$).

It is now obvious that the PSR define in (\ref{eq:mPSR}) generalizes to situations outside of the usual uncensored and left, right, and interval censored situations. 
\end{comment}

The PSR for this general setup is defined as
\begin{equation}
\label{eq:uPSR}
  r(y,F^*)=
  \begin{cases}
      F^*(t)+F^*(t-)-1, & \text{if $y$ is an event time $t$}; \\
      F^*(C_j)+F^*(C_{j+1})-1, & \text{if $y$ is an interval $(C_j,C_{j+1}]$}.
  \end{cases}
\end{equation}
%\begin{equation}
%\label{eq:uPSR}
%  R= \sum_{j=0}^{K} I_{(C_j < T\le C_{j+1})} \{ \pi_j(F^*(T)+F^*(T-)-1) \oplus (1-\pi_j)(F^*(C_j)+F^*(C_{j+1})-1) \},
%\end{equation}
%where $\{\oplus\}$ indicates a mixture distribution: the value between the curly brackets is $F(T)+F(T-)-1$ with probability $\pi_j$ and $F(C_j)+F(C_{j+1})-1$ with probability $1-\pi_j$. 
This expression covers all previously defined PSR. For purely interval-censored data (S6), (\ref{eq:uPSR}) becomes (\ref{eq:PSR}) since $\pi_j\equiv0$. For data with no censoring (S1), (\ref{eq:uPSR}) becomes (\ref{eq:psr}) since $\pi_j\equiv1$. For right-censored data (S2), when the outcome is censored (i.e., $y$ is $(C,\infty)$), the PSR becomes $F^*(C)$, as was shown in \cite{shepherd2016probability}. For current status data (S5), the PSR is $F^*(C)$ when the outcome is $(C,\infty)$, and $F^*(C)-1$ when the outcome is $(0,C)$ \citep{shepherd2016probability}.

%and $I_{(C_j < T\le C_{j+1})} \{(F^*(C_j)+F^*(C_{j+1})-1)\} = F^*(U) + F^*(L) - 1$, when $T \in (L, U]$ where $L$ and $U$ are defined in Section \ref{sec:method} and 0 otherwise. Lastly, we assume $T \perp (C, K, \pibf)$. In the regression setting, we assume independence conditional on the covariates included in the model.\\

%The practical application of this PSR is that, regardless of the subject's censoring status, it can be calculated using the same general formula, $F^*(u) + F^*(l) - 1$, where $l$ and $u$ are defined accordingly. 

We now state the main results for the PSR is this general setup. Let $R$ be the PSR defined in (\ref{eq:uPSR}) but is viewed as a random variable (i.e., a function of the random outcome $Y$). Then we have
\begin{theorem} When $F^*=F$ and $T \perp (K, \Cbf, \pibf)$,
\begin{enumerate}
    \item[P1$^*$)] $E(R) = 0$.
    \item[P2$^*$)] $Var(R) = E_{K,\Cbf,\pibf} [E(R^2\mid K,\Cbf,\pibf)]$, where the conditional variance is
    $$E(R^2\mid K,\Cbf,\pibf) =\sum_{j=0}^k (\pi_j p_j^3 v_j(K,\Cbf) + p_j r_j^2),$$ with\, 
    $p_j = F(C_{j+1}) - F(C_j)$, $r_j = F(C_{j+1}) + F(C_j) - 1$, and
    \begin{align*}
        v_j(K,\Cbf) = \frac{1}{p_j^3} \int_{(C_j, C_{j+1}]} \left(F(t)+F(t-) - 1 - r_j\right)^2dF(t).
    \end{align*}  
\end{enumerate}
\end{theorem}
The proofs are in the Appendix. We now simplify the conditional variance of the PSR, $E(R^2\mid K,\Cbf,\pibf)$, for each of the six special cases we mentioned above. 

\begin{enumerate}
    \item[S1.] Continuous event-time data with no censoring: $\pi_j\equiv 1$, $v_j(K,\Cbf)\equiv 1/3$ for all $j$, and
\begin{equation*}
 E(R^2\mid K,\Cbf,\pibf) =\frac{1}{3} \sum_{j=0}^{k} \left( p_j^3 + 3p_jr_j^2 \right) =\frac{1}{3}.
\end{equation*}
    \item[S2.] Right-censored data: $K=1$, $\Cbf=C$ is the single censoring variable, $\pibf=(1,0)$, and
\begin{align*}
  E(R^2\mid K,\Cbf,\pibf) &=v_0(1,C)F(C)^3 +F(C)(F(C)-1)^2 +(1-F(C))F(C)^2 \\
  &= v_0(1,C)F(C)^3 + F(C)(1 - F(C)),
\end{align*}
where $v_0(1,C)= \frac{1}{F(C)^3} \int_{(0, C]} \left(F(t)+F(t-) -F(C)\right)^2dF(t)$.
    \item[S3.] Left censored data: $K=1$, $\Cbf=C$ is the single censoring variable, $\pibf=(0,1)$, and
\begin{align*}
  E(R^2\mid K,\Cbf,\pibf)
  &= F(C)(F(C)-1)^2 +v_1(1,C)(1-F(C))^3 +(1-F(C))F(C)^2 \\
  &= v_1(1,C)(1-F(C))^3 + F(C)(1-F(C)),
\end{align*}
where $v_1(1,C)= \frac{1}{(1-F(C))^3} \int_{(C,\infty)} \left(F(t)+F(t-) -1-F(C)\right)^2dF(t)$.
    \item[S4.] Doubly censored data: $K=2$, $\Cbf=(C_1,C_2)$, $\pibf=(0,1,0)$, and
\begin{align*}
  &\ E(R^2\mid K,\Cbf,\pibf) \\
  =&\ v_1(2,\Cbf)(F(C_2)-F(C_1))^3 + F(C_1)^3 - 3F(C_1)^2 + 2F(C_1) - F(C_2)^3 +2F(C_2)^2 - F(C_2),
\end{align*}
where $v_1(2,\Cbf)= \frac{1}{(F(C_2)-F(C_1))^3} \int_{(C_1,C_2]} \left(F(t)+F(t-) -F(C_1)-F(C_2)\right)^2dF(t)$.
    \item[S5.] Current status data: $K=1$, $\Cbf=C$ is the single censoring variable, $\pibf=(0,0)$, and
\begin{align*}
 E(R^2\mid K,\Cbf,\pibf)
 &= F(C)(F(C) - 1)^2 +(1-F(C))F(C)^2 \\
 &= F(C)(1-F(C)).
\end{align*}
    \item[S6.] interval-censored data: $\pi_j\equiv 0$, and
\begin{align*}
  E(R^2\mid K,\Cbf,\pibf)
  &= \sum_{j=0}^{k} (F(C_{j+1}) - F(C_j))(F(C_{j+1}) + F(C_j)-1)^2 \\
  &= \sum_{j=1}^{k} F(C_{j+1})F(C_j) [F(C_{j+1}) - F(C_j)].
\end{align*}
\end{enumerate}
Note that the conditional variance for the first two settings coincide with the formulas provided by \cite{shepherd2016probability} \blacktext{for continuous and right-censored data}.  The conditional variance for interval-censored data (S6) agrees with what we derived in Section \ref{sec:method}.  For current status data (S5), \cite{shepherd2016probability} showed that the PSR is $F^*(c) - 1$ when outcome is left censored at $c$ and is $F^*(c)$ when outcome is right censored at $c$; but did not calculate the variance of the PSR. From Theorem 1, the variance for the current status data PSR is $E_C [F(C)(1 - F(C))]$.

\subsection{PSR under equally-spaced inspection times and exponentially distributed event times}

We now consider a simple case $k$ interval censoring setting with fixed inspection times. Let $T$ be exponentially distributed with rate parameter $\lambda$ and assume we have equally spaced inspection times $(0, \tau, 2\tau, 3\tau, \ldots)$ for some $\tau > 0$. Then $T \in (K\tau, (K+1)\tau]$ for some $K = \lfloor T/\tau \rfloor$. The PSR, under the true distribution $F$, is
\begin{align}
    r\{(k\tau, (k+1)\tau], F\} & = F((k+1)\tau) + F(k\tau) - 1 \notag \\
    & = 1 - \exp(-\lambda\tau)^k\{1 + \exp(-\lambda \tau)\}.
\end{align}

%A natural question to ask is how the PSR is affected by $\tau$, the width of the interval.
As the interval shrinks, one expects the left and right endpoints of the interval to converge to the true event time. Formally, $k\tau \to T^-$ and $(k+1)\tau \to T^+$ when $\tau \to 0$. Therefore, as $\tau \to 0$, $r\{(k\tau, (k+1)\tau], F\} \to 1 - \exp(-\lambda T) - \exp(-\lambda T) = 1 - 2\exp(-\lambda T) \equiv 2F(t) - 1$, the PSR for continuous outcomes \citep{shepherd2016probability}. %On the other hand, we have very little information about the event time when the interval widens. Since $k = \lfloor T/\tau \rfloor$, $k = 0$ whenever $T < \tau$. Thus, $k\tau \downarrow 0$ and $(k+1)\tau \uparrow \infty$ for $T < \tau \to \infty$ and $r\{(k\tau, (k+1)\tau], F\} \to 0$. In addition, we show that the PSR under Case I interval censoring becomes the right and left-censored versions of the PSR by moving the right and left endpoints, respectively for suitably small $\tau$. Note that for any $x \in \mathbb{R}$, $\left\| x/\tau - \lfloor x/\tau \rfloor \right\| \to 0$  as $\tau \to 0$. If $T \in (k\tau, (k+1)\tau]$, then $T \in (k\tau, u^*\tau]$ for any $u^* \geq (k + 1)$. As $u^* \to \infty$ and $\tau \to 0$, $r\{(k\tau, u^*\tau], F\} \to 1 - \exp(-\lambda k\tau) \approx 1 - \exp(-\lambda T) \equiv F(t)$. Conversely, if $T \in (k\tau, (k+1)\tau]$, then $T \in (l^*\tau, (k+1)\tau]$ for any $l^* < k$ and $r\{(l^*\tau, (k+1)\tau], F\} \to -\exp(-\lambda (k + 1)\tau) \approx -\exp(-\lambda T) \equiv F(t) - 1$ as $l^* \to 0$ and $\tau \to 0$.
We can also derive the sampling distribution of the PSR under this scenario. For $x \in (-1, 1)$, the cumulative distribution function of the PSR, $R$, is given as
\begin{align*}
    \Pr(R < x) = 1 - \exp(-\lambda \tau)^{-\lfloor \frac{1}{\lambda \tau} \log \left(\frac{1-x}{1 + \exp(-\lambda \tau)}\right)\rfloor + 1}.
\end{align*}
%Details are provided in the Appendix.  
As $\lambda\tau \to 0$,
\begin{align}
    \Pr(R < x) \to 1 - \left(\frac{1-x}{2}\right) = \left(\frac{x + 1}{2}\right),
\end{align}
which is the cumulative distribution function of $\mathcal{U}(-1, 1)$. This is expected since $\tau \to 0$ suggests that the interval containing the true event time $T$ is 
 converging to $T$. %In addition, as $\tau \to \infty$,  $\Pr(r < x) \to 1$ and, under mild assumptions, the sampling density for the PSR converges to $\mathcal{U}[0, 1]$ when $u \to \infty$ (with $l$ fixed) and $\mathcal{U}[-1, 0]$ when $l \to 0$ (with $u$ fixed).

\section{Application to CCASAnet Data}
\label{sec:rda}
We demonstrate the use of the PSR with data from the Carribean, Central and South America network (CCASAnet) for HIV epidemiology. 
Data were collected on people living with HIV at one of seven sites in Latin America and the Caribbean \citep{mcgowan2007cohort}. For illustrative purposes, we focus our attention on women who were $\geq 50$ years of age at initiation of antiretroviral therapy (ART) and have not yet observed an AIDS defining event (ADE). 

The purpose of this study was to understand ADE-free survival after ART initiation. ADEs occur between two consecutive follow-up times and are thus interval censored \citep{gao2017semiparametric}. Date of death, however, can be obtained through public records and is exactly observed. Subjects who are alive and ADE-free by the end of the follow-up period are right censored. Such data are commonly referred to as partly-interval censored (PIC) since some of the failure times are exactly observed, while other failure times are subject to interval censoring. Of the 1380 subjects in our study, 143 had died without an ADE during follow-up, 185 individuals had an ADE, the remaining 1052 were right-censored (ADE free and alive) at last follow-up. A Weibull parametric accelerated failure time model was fit using the R package \verb+icensReg+ \citep{anderson2017icenreg}.  The site-stratified model included age at ART initiation, biological sex and ART class (HAART vs. Non-HAART) as covariates in the model. By the definition provided in Section \ref{sec:upsr}, the PSR is
\[ PSR_i = \begin{cases} 
      2F^*(y_i) - 1 & \mbox{if subject $i$ died at time $y_i$ without an ADE;} \\
      F^*(u_i) + F^*(l_i) - 1 & \mbox{if subject $i$ had an ADE between ($l_i$, $u_i$];} \\
    F^*(c_i) & \mbox{if subject $i$ was right censored at time $c_i$.} 
   \end{cases}
\]

\begin{figure}
    \centering
    \includegraphics[scale = 0.35]{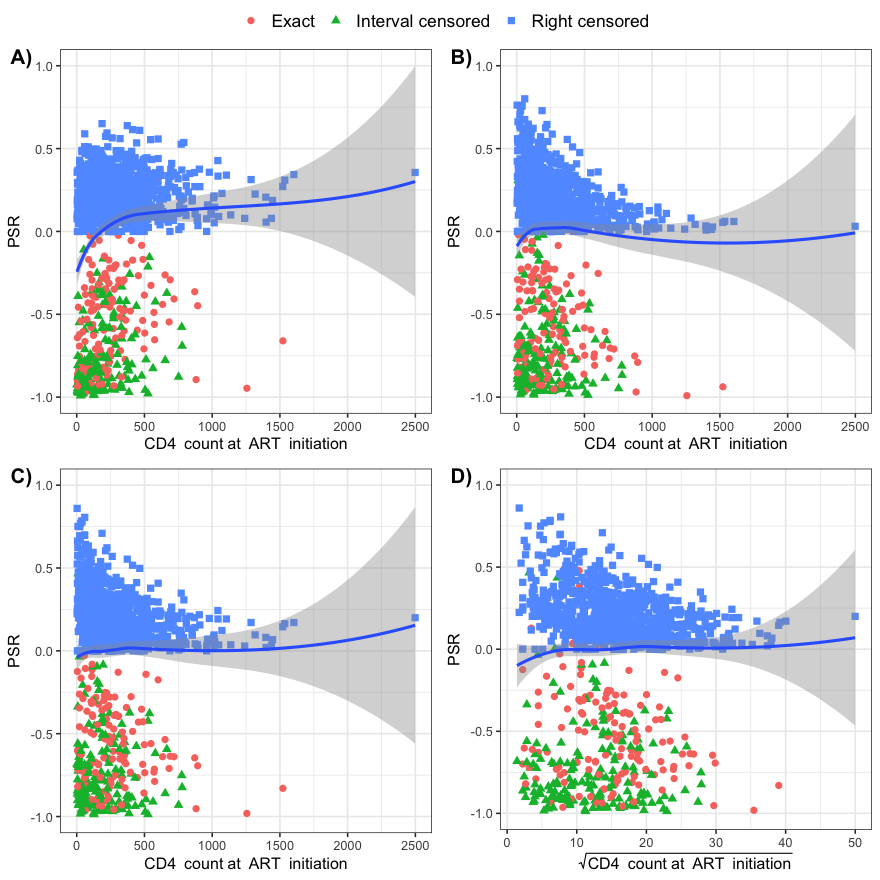}
\caption{Plots of PSR vs. CD4 count at ART initiation.  A) Plot for a model that does not include CD4 count at ART initiation; B) Plot for a model that includes ART initiation after square root transformation; C) Plot for a model that models square root ART initiation as a piece-wise linear function with a break point at 18; D) Plot of PSR vs. square root of CD4 count at ART initiation using the same model as in Panel C.  Exact events (i.e., ADE-free death) are denoted with circles, interval censoring (i.e., ADE) are denoted with triangles and right censoring (ADE-free and alive) are denoted with squares.}
\label{fig:cd4}
\end{figure}

\begin{figure}
    \centering
    \includegraphics[scale = 0.35]{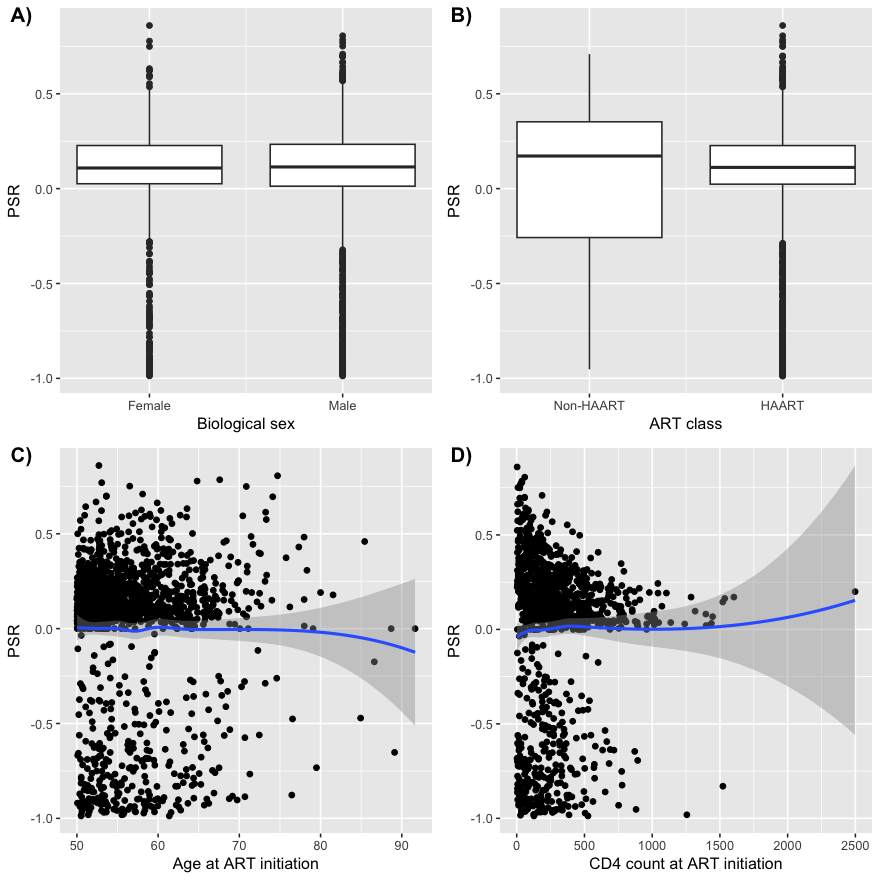}
\caption{Plots of PSR vs. covariates in the final model.  A) Biological sex; B) ART class; C) Age at ART initiation; D) Square root of CD4 count at ART initiation using a natural spline.  Exact events (i.e., ADE-free death) are denoted with circles, interval censoring (i.e., ADE) are denoted with triangles and right censoring (ADE-free and alive) are denoted with squares.}
\label{fig:covplot}
\end{figure}

\begin{figure}
    \centering
    \includegraphics[scale = 0.4]{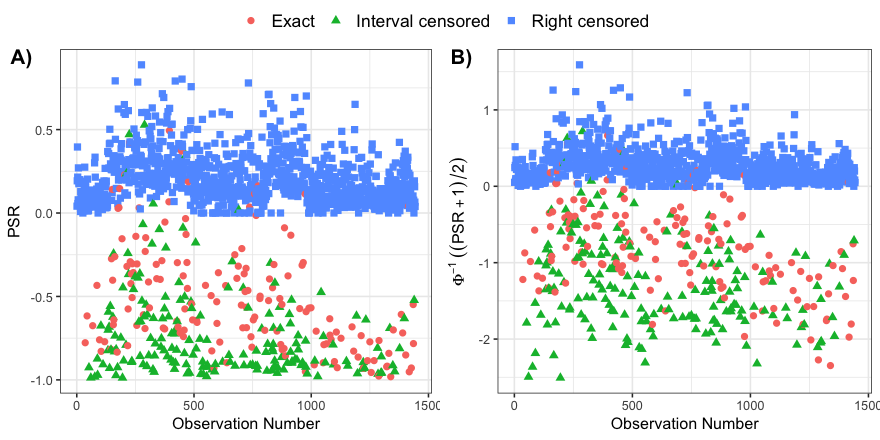}
\caption{PSR for CCASAnet example. A) Plot of PSR by observation number; B) Plot of transformed PSR by observation number. Exact events (i.e., ADE-free death) are denoted with circles, interval censoring (i.e., ADE) are denoted with triangles and right censoring (ADE-free and alive) are denoted with squares.}
\label{fig:outlier}
\end{figure}

Panel A in Figure \ref{fig:cd4} shows PSRs from this model plotted against CD4 count at ART initiation, a covariate that is known to be associated with both ADE and mortality but not included in the model. As expected, the PSRs for right-censored patients (blue squares) are $\geq 0$. The PSRs for those who died (red circles) or experienced an ADE (green triangles) are $<0$, although this need not be the case. Since the PSR should, on average, have mean zero, departures from a slope of zero may indicate a need to include the covariate or transform the covariate if it is included in the model. The smoothed locally estimated scatterplot smoothing (LOESS) curve suggests both a non-zero and non-linear relationship between CD4 count and ADE/mortality.  Panel B show residuals from the model with CD4 included after a square-root transformation. The relationship between CD4 and the PSR shows improvement; however, there is some curvature at lower values of CD4 count and PSRs from the model without CD4 count (Panel A), suggesting CD4 count should be included in the model, possible in a non-linear manner. Upon adjusting for non-linearity with a natural spline with 3 knots, the relationship between CD4 count and the PSR diminishes (Panel C). This is further corroborated in Panel D, where we plot the PSR against the square root of CD4 count. 

In addition, we also present PSR-vs-predictor plots for the other three variables we included in our updated model (Figure \ref{fig:covplot}). No obvious relationships between PSRs and covariates were observed for biological sex, ART class, and age at ART initiation. Lastly, we present an index plot that plots the PSR against observation number (Figure \ref{fig:outlier}) using the model that includes the square root transformed CD4 count with a natuaral spline. This type of plot may be helpful to identify outliers.  Since the range of the PSR is restricted (Panel A), we apply the following transformation to the PSR $f(x) = \Phi^{-1}((x+1)/2)$, where $\Phi^{-1}(\cdot)$ is the inverse of the cumulative distribution function of the standard normal distribution (Panel B). While some observations have somewhat extreme values for the PSR ($<-2$), this is not unusual as we have 1380 individuals in our study. 

\section{Conclusion}
\label{sec:disc}
We have extended the PSR to a wide range of time-to-event settings including mixed case interval-censored data. Not only is the probability-scale residual (PSR) applicable in a wide range of scenarios but, as as shown in Section \ref{sec:upsr}, it can also be written using the same definition. This feature of the PSR is quite remarkable as the residual definition need not be outcome specific. 

However, the generalizability of the PSR also poses some limitations \cite{shepherd2016probability}. While symmetric in range, the PSR is not ideal for outlier detection since it is bounded between -1 and 1. To remedy this, \cite{shepherd2016probability} suggested the following transformation, $\Phi^{-1}\{(PSR + 1)/2\},$ where $\Phi^{-1}(\cdot)$ is the inverse of the CDF of the standard normal distribution, as demonstrated in Section \ref{sec:rda}. This transformation aims to magnify the extreme values of the PSR, improving its ability to identify outliers. The PSR may also provide little or no information on the adequacy of some model assumptions. For example, Schoenfeld residuals are often used to test the proportional hazards assumption. While the PSR have several advantages, they may not be strong enough to persuade analysts to switch to the PSR for their {\em{primary}} model diagnostic. Although ideal, no single residual can be uniformly superior in all cases. One appeal of the PSR is its usefulness across a wide range of outcome types using the same definition.

Although the PSR proposed in this paper focuses on general censoring mechanisms for time-to-event data, its application is not limited to this context. For example, with detection limit data, censoring occurs when a value falls beyond the measuring instrument's range. In this case $\Cbf$ may not be a vector of inspection times but instead a vector (or in most cases a scalar) of detection limits. Lastly, while our extension of the PSR encompasses several censoring schemes,  more complex situations have yet to be explored.  In our real data example, we defined the event of interest as ADE-free survival. If instead, our primary interest was solely modeling time to an ADE, which is interval censored, then ADE-free mortality, which is exactly observed, would be considered a competing risk.  Developing the PSR for more complex time-to-event settings is an avenue of research that the authors are currently exploring. 

\section*{Acknowledgements}
The authors thank the Caribbean, Central, and South American Network (CCSAnet) for HIV Epidemiology for the use of their data. This work was supported in part by U.S. National Institutes of Health grants U01AI069923 and R01AI093234. 

\newpage
\section*{Appendix}

\subsection*{Derivation of equation (\ref{eq:PSR}), $r\{(l,u], F^*\}= F^*(u)+F^*(l)-1$}

The PSR, $r\{(l,u], F^*\}$, is the conditional expectation of $r(T^*, F^*)$ with respect to $T ^*\sim F^*$ conditional on $T^*\in (l, u]\}$.  Let $G(t)$ be the CDF of the conditional distribution of $T^*$ given $T^*\in(l,u]$.  Then $G(t)=\frac{F^*(t)-F^*(l)}{F^*(u)-F^*(l)}$ for $t\in(l,u]$, and
\begin{align*}
    r\{(l, u], F^*\} 
    &= \int_{(l,u]} \{F^*(t)+F^*(t-)-1\} dG(t) \\
    &= \frac{1}{F^*(u)-F^*(l)} \int_{(l,u]} \{F^*(t)-F^*(l) +F^*(t-)-F^*(l) +2F^*(l)-1\} dF^*(t).
\end{align*}
Since
\begin{align*}
    \int_{(l,u]} \{F^*(t)-F^*(l)\} dF^*(t) 
    &= \int_{t\in(l,u]} \int_{s\in(l,t]} dF^*(s) dF^*(t), \\
    \int_{(l,u]} \{F^*(t-)-F^*(l)\} dF^*(t) 
    &= \int_{t\in(l,u]} \int_{s\in(l,t)} dF^*(s) dF^*(t) =\int_{s\in(l,u)} \int_{t\in(s,u]} dF^*(s) dF^*(t) \\
    &= \int_{t\in(l,u)} \int_{s\in(t,u]} dF^*(s) dF^*(t),
\end{align*}
these two terms add up to be
\begin{align*}
    \int_{t\in(l,u]} \int_{s\in(l,u]} dF^*(s) dF^*(t) =\{F^*(u)-F^*(l)\}^2.
\end{align*}
Then
\begin{align*}
    r\{(l, u], F^*\} 
    &= \frac{\{F^*(u)-F^*(l)\}^2 +\{2F^*(l)-1\}\{F^*(u)-F^*(l)\}}{F^*(u)-F^*(l)} \\
    &= F^*(u)+F^*(l)-1.
\end{align*}

\begin{comment}
\begin{align*}
    r\{(l, u], F^*\} & = E_{T} \{ r(T, F^*) | T \in (l, u]\} \\
    & = E_{T} \{ r(T, F^*) | L = \max\{C_k: C_k < T\}, U = \min\{C_k: C_k \geq T\}, L, U\} \\
    & = 
    \frac{E_{T} \left[ r(T, F^*) \times I(L = \max\{C_k: C_k < T\}, U = \min\{C_k: C_k \geq T\}) | C_{1}, \ldots, C_{K}, K \right]}{\Pr(L = \max\{C_k: C_k < T\}, U = \min\{C_k: C_k \geq T\}) | C_{1}, \ldots, C_{K}, K )} \\
    & = 
    \frac{E_{T} \left[ \sum_{k=1}^K r(T, F^*) \times I(C_k = L, C_{k + 1} = U, L < T \leq U) | C_{1}, \ldots, C_{K}, K\right]}{\sum_{k=1}^K\Pr(C_k = L, C_{k + 1} = U, L < T \leq U | C_{1}, \ldots, C_{K}, K)} \\
    & = 
    \frac{E_{T} \left[ r(T, F^*) \times I(L < T \leq U) | L, U \right] \times 
    E\left[ \sum_{k=1}^K I(L = C_k, U = C_{k+1} | L, U)\right]
    }{\Pr(L < T \leq U | L, U) \times E\left[ \sum_{k=1}^K I(L = C_k, U = C_{k+1} | L, U)\right]} \\
     & = 
    \frac{E_{T} \left[ r(T, F^*) \times I(L < T \leq U) | L, U \right]}{\Pr(L < T \leq U | L, U)} \\
    & =  \frac{\int_{t \in (l, u]} r(T, F^*)dF^*(t)}{F^*(U) - F^*(L)} \\
    & = \frac{\int_{t > L} r(T, F^*)dF^*(t) - \int_{t \geq U} r(T, F^*)dF^*(t)}{F^*(U) - F^*(L)} \\
    & = \frac{F^*(L)[1 - F^*(L)] - F^*(U)[1 - F^*(U)]}{F^*(U) - F^*(L)} \\
    & = F^*(U) + F^*(L) - 1,
\end{align*}
where the fifth equality is due to the independence of $(K, \Cbf)$ and $T$ and the ninth equaity is due to \cite{shepherd2016probability}.
\end{comment}

\subsection*{Derivation of P4 and P5, the Expectation and Variance of the PSR}

Since $T\perp (K,\Cbf)$ and random variables $L$ and $U$ are functions of $(K,\Cbf)$, we have $T\perp (L,U)$. Given fixed values of $K$ and $\Cbf=(C_1,\ldots,C_K)$, let $f_k=F(C_k)$ for $k=0,\ldots,K+1$.  Then $f_0=F(0)=0$ and $f_{K+1}=F(\infty)=1$. For brevity, let $R = r\{(L, U], F\} =F(L)+F(U)-1$. Then
\begin{align*}
    E\{R\mid (K,\Cbf)\} & = \sum_{k=0}^K (f_{k+1} +f_k-1)(f_{k+1}-f_k) \\
    & = \sum_{k=0}^K (f_{k+1}^2 -f_k^2) - \sum_{k=0}^K (f_{k+1}-f_k) \\
    & = f_{K+1}^2 -f_{K+1} \\
    & = 0.
\end{align*}
Thus, $E(R) = E_{(K,\Cbf)}[E\{R\mid (K,\Cbf)\}] = 0$.  The variance is $Var(R)= E(R^2)=E_{(K,\Cbf)} \{R^2\mid (K,\Cbf)\}$, where
\begin{align*}
    E_{(K,\Cbf)}\{R^2\mid (K,\Cbf)\}
    &= \sum_{k=0}^K (f_{k+1} + f_k-1)^2 (f_{k+1}-f_k) \\
    &= \sum_{k=0}^K \left[ (f_{k+1}+f_k)^2 - 2(f_{k+1}+f_k) + 1 \right](f_{k+1}-f_k) \\
    & = \sum_{k=0}^K  (f_{k+1}+f_k)^2 (f_{k+1}-f_k)
    -2\sum_{k=0}^K (f_{k+1}^2-f_k^2) +\sum_{k=0}^K (f_{k+1}-f_k) \\
    & = \sum_{k=0}^K (f_{k+1}+f_k)^2(f_{k+1}- f_k) - 1.
\end{align*}
Expanding the first term further, we have
\begin{align*}
    \sum_{k=0}^K (f_{k+1}+f_k)^2 (f_{k+1}-f_k)
    &= \sum_{k=0}^K (f_{k+1}^2+f_k^2  +2f_{k+1}f_k) (f_{k+1}-f_k) \\
    &= \sum_{k=0}^K (f_{k+1}^3-f_k^3) + \sum_{k=0}^K f_{k+1}f_k(f_{k+1} - f_k) \\
    &= 1 + \sum_{k=1}^K f_{k+1}f_k(f_{k+1} - f_k).
\end{align*}
Thus,
\begin{align*}
    E[R^2|(K,\Cbf)] & = \sum_{k=1}^K f_{k+1}f_k(f_{k+1} - f_k) \\
    &=\sum_{k=1}^K F(C_{k+1})F(C_k)[F(C_{k+1}) - F(C_k)].
\end{align*}

\subsection*{Derivation of the sampling distribution of the PSR under a simple scenario.}

Recall that $T$ is exponentially distributed with rate parameter $\lambda$ and assume we have equally spaced inspection times $(0, \tau, 2\tau, 3\tau, \ldots)$ for some $\tau > 0$. Then, $T \in (k\tau, (k+1)\tau]$ where $k = \lfloor T/\tau \rfloor$. The PSR, under the true distribution $F$, is
\begin{align}
    r(t, F) & = F((k+1)\tau) + F(k\tau) - 1 \notag \\
    & = \left(\exp(-\lambda\tau)\right)^k\{1 - \exp(-\lambda \tau)\}.
\end{align}

Since $T$ is exponentially distributed with rate $\lambda$ and $k = \lfloor T/\tau \rfloor$, 1) $T/\tau \sim \mathcal{E}(\lambda \tau)$ and 2) $k$ is geometric with probability $1 - \exp(-\lambda \tau)$. For brevity, let $r = r\{(l, u], F)$. Then for any $x \in (-1, 1)$,
\begin{align*}
    \Pr(R \leq x) & = \Pr\left(\exp(-\lambda\tau)^k\{1 - \exp(-\lambda \tau)\} \leq x\right) \\
    & = \Pr(k \leq u) \\
    & = 1 - \exp(-\lambda \tau)^{\lfloor u \rfloor + 1} \\
    & = 1 - \exp(-\lambda \tau)^{-\lfloor \frac{1}{\lambda \tau} \log \left(\frac{1-x}{1 + \exp(-\lambda \tau)}\right)\rfloor + 1}
\end{align*}

\subsection*{The PSR as expected values under the $\mathcal{U}(-1, 1)$ distribution.}

As with the Cox-Snell residual, naively calculating the PSR at the lower and upper endpoint of an interval will lead to a residual that is also an interval. One way we can avoid this problem will be to replace the interval residuals with expected values under a $\mathcal{U}(-1, 1)$ distribution. \\

Let $R \sim \mathcal{U}(-1, 1)$.   Then,
\begin{align*}
    E\{R|R \in (2F^*(l) - 1, 2F^*(u) - 1]\} & = \frac{\int_{2F^*(l)-1}^{2F^*(u)-1}r \frac{1}{2}dr}{\int_{2F^*(l)-1}^{2F^*(u)-1} \frac{1}{2}dr} \\
    & = \frac{(F^*(u) - F^*(l))(F^*(u)+F^*(l)-1)}{F^*(u)-F^*(l)} \\
    & = F^*(u)+F^*(l)-1.
\end{align*}
Thus, the PSR can be viewed as expected value of the interval residuals under the $\mathcal{U}(-1, 1).$ Furthermore, the PSR can also equivalently be defined by taking the average of the PSR for continuous outcomes evaluated at both the lower and upper endpoints:
\begin{align*}
    \frac{r(u, F^*) + r(l, F^*)}{2} = \frac{\{2F^*(u) - 1\} + \{2F^*(l) - 1\}}{2} & = F^*(u) + F^*(l) - 1.
\end{align*}

\subsection*{Derivation of P1$^*$ and P2$^*$, the mean and variance of the PSR in Section \ref{sec:upsr}}

We will show that $E(R)=0$, and derive $Var(R)$ under the assumption of independence between $T$ and $(K,\Cbf,\pi)$.  Given $(K,\Cbf)$, let $f_k=F(C_k)$ for $k=0,\ldots,K+1$; then $f_0=0$ and $f_{K+1}=1$.  

\begin{lemma}
For any $j = 0,1,\ldots,K$, let $v_j(K,\Cbf)$ be the variance of the PSR for the conditional distribution of $T$ given $C_j<T\le C_{j+1}$.  We have the following results:
\begin{align*}
  & E[I_{(C_j<T\le C_{j+1})} (F(T)+F(T-)-f_j-f_{j+1}) \mid K,\Cbf] =0; \\
  & E[I_{(C_j<T\le C_{j+1})} (F(T)+F(T-)-1) \mid K,\Cbf] =(f_{j+1}-f_j)(f_j+f_{j+1}-1); \\
  & E[I_{(C_j<T\le C_{j+1})} (F(T)+F(T-)-1)^2 \mid K,\Cbf]  = v_j(K,\Cbf)(f_{j+1}-f_j)^3 +(f_{j+1}-f_j)(f_j+f_{j+1}-1)^2.
\end{align*}
\end{lemma}

\begin{proof}[Proof of Lemma 1]
 We give detailed derivation of the three equations for $k=1$; the derivation for the other $k$'s is similar.  Consider the conditional distribution of $T$ given $C_1<T\le C_2$.  Its support is $(C_1,C_2]$ and its CDF over this interval is $G(t)=\frac{F(t)-F(C_1)}{F(C_2)-F(C_1)}=\frac{F(t)-f_1}{f_2-f_1}$.  We know the PSR for distribution $G$ has mean 0; that is,
\begin{align*}
  0 &=\int_{(C_1,C_2]} (G(t)+G(t-)-1)\ dG(t) \\
  &=\int_{(C_1,C_2]} \frac{(F(t)-f_1)+(F(t-)-f_1)-(f_2-f_1)}{f_2-f_1}\ d \frac{F(t)-f_1}{f_2-f_1} \\
  &=\frac{1}{(f_2-f_1)^2} \int_{(C_1,C_2]} (F(t)+F(t-)-f_1-f_2)\ dF(t) \\
  &=\frac{1}{(f_2-f_1)^2} E[I_{(C_1<T\le C_2)} (F(T)+F(T-)-f_1-f_2)].
\end{align*}
Then
\begin{align*}
  &E[I_{(C_1<T\le C_2)} (F(T)+F(T-)-f_1-f_2)] =0, \\
  &E[I_{(C_1<T\le C_2)} (F(T)+F(T-)-1)] =E[I_{(C_1<T\le C_2)} (f_1+f_2-1)] =(f_2-f_1)(f_1+f_2-1).
\end{align*}

Because the PSR for distribution $G$ has mean 0, its variance is
\begin{align*}
  v_1(K,C) &=\int_{(C_1,C_2]} (G(t)+G(t-)-1)^2\ dG(t) \\
  &=\int_{(C_1,C_2]} \frac{[(F(t)-f_1)+(F(t-)-f_1)-(f_2-f_1)]^2}{(f_2-f_1)^2}\ d \frac{F(t)-f_1}{f_2-f_1} \\
  &=\frac{1}{(f_2-f_1)^3} \int_{(C_1,C_2]} (F(t)+F(t-)-f_1-f_2)^2\ dF(t).
\end{align*}
Then $E[I_{(C_1<T\le C_2)} (F(T)+F(T-)-f_1-f_2)^2 | K,C] = v_1(K,C)(f_2-f_1)^3$, and
\begin{align*}
  & E[I_{(C_1<T\le C_2)} (F(T)+F(T-)-1)^2 | K,C] \\
  =& E[I_{(C_1<T\le C_2)} (F(T)+F(T-)-f_1-f_2+f_1+f_2-1)^2 | K,C] \\
  =& E[I_{(C_1<T\le C_2)} (F(T)+F(T-)-f_1-f_2)^2 | K,C] \\
  & +E[I_{(C_1<T\le C_2)} 2(F(T)+F(T-)-f_1-f_2)(f_1+f_2-1) | K,C] \\
  & +E[I_{(C_1<T\le C_2)} (f_1+f_2-1)^2 | K,C] \\
  =& v_1(K,C)(f_2-f_1)^3 +(f_2-f_1)(f_1+f_2-1)^2.
\end{align*}
\end{proof}

{\em{Mean and variance of the unified PSR}}

To show $E(R)=0$, it suffices to show $E(R|K,C,\pi)=0$:
\begin{align*}
  E(R|K,C,\pi)
  &= \sum_{j=0}^{k} \{ \pi_j (f_{j+1}-f_j)(f_j+f_{j+1}-1) + (1-\pi_j) (f_{j+1}-f_j)(f_j+f_{j+1}-1) \} \\
  &= \sum_{j=0}^{k} (f_{j+1}-f_j)(f_j+f_{j+1}-1) =\sum_{j=0}^{k} [(f_{j+1}^2-f_j^2)-(f_{j+1}-f_j)] = 1-1=0.
\end{align*}

The conditional expectation of $R^2$ given $(K,C,\pi)$ is
\begin{align}
  &\sum_{j=0}^{k} \left( \pi_j [v_j(K,C)(f_{j+1}-f_j)^3 +(f_{j+1}-f_j)(f_j+f_{j+1}-1)^2] + (1-\pi_j)(f_{j+1}-f_j)(f_j+f_{j+1}-1)^2 \right) \nonumber \\
  =&\sum_{j=0}^{k} \left( \pi_j v_j(K,C)(f_{j+1}-f_j)^3 + (f_{j+1}-f_j)(f_j+f_{j+1}-1)^2 \right).
  \label{ER2}
\end{align}
The variance of $R$ is the expectation of (\ref{ER2}) over the distribution of $(K,C,\pi)$.  When $T$ is absolutely continuous, $v_j(K,C)\equiv 1/3$, and (\ref{ER2}) becomes
\begin{equation*}
  \sum_{j=0}^{k} \left( \frac{\pi_j}{3} (f_{j+1}-f_j)^3 + (f_{j+1}-f_j)(f_j+f_{j+1}-1)^2 \right).
\end{equation*}

\bibliographystyle{abbrvnat}
\bibliography{references.bib}

\end{document}